\documentclass
[superscriptaddress,secnumarabic,amssymb,amsmath,nobibnotes,aps,prd,showkeys,showpacs,nofootinbib,onecolumn,notitlepage,10pt]{revtex4}%
\usepackage{setspace}
\usepackage{color}
\usepackage{amsmath}
\usepackage{amsfonts}
\usepackage{verbatim}
\usepackage{amssymb}
\usepackage{graphicx,bm}
\usepackage{amsmath}
\usepackage{amssymb}
\usepackage{graphicx}
\usepackage{pdflscape}
\usepackage{subcaption}
\usepackage[caption=false]{subfig}%
\providecommand{\U}[1]{\protect\rule{.1in}{.1in}}
\newcommand{\be}{\begin{equation}}
\newcommand{\ee}{\end{equation}}

\newcommand{\mincir}{\raise
-3.truept\hbox{\rlap{\hbox{$\sim$}}\raise4.truept\hbox{$<$}\ }}
\newcommand{\magcir}{\raise
-3.truept\hbox{\rlap{\hbox{$\sim$}}\raise4.truept\hbox{$>$}\ }}

\ifx\pdfoutput\relax\let\pdfoutput=\undefined\fi
\newcount\msipdfoutput
\ifx\pdfoutput\undefined\else
\ifcase\pdfoutput\else
\ifx\paperwidth\undefined\else
\ifdim\paperheight=0pt\relax\else\pdfpageheight\paperheight\fi
\ifdim\paperwidth=0pt\relax\else\pdfpagewidth\paperwidth\fi
\fi\fi\fi
\begin{document}
\title{Observational Constraints on Dark Energy Models with $\Lambda$ as an
Equilibrium Point}
\author{Andronikos Paliathanasis}
\email{anpaliat@phys.uoa.gr}
\affiliation{Department of Mathematics, Faculty of Applied Sciences, Durban University of
Technology, Durban 4000, South Africa}
\affiliation{Centre for Space Research, North-West University, Potchefstroom 2520, South Africa}
\affiliation{Departamento de Matem\`{a}ticas, Universidad Cat\`{o}lica del Norte, Avda.
Angamos 0610, Casilla 1280 Antofagasta, Chile}
\affiliation{National Institute for Theoretical and Computational Sciences (NITheCS), South Africa.}

\begin{abstract}
We investigate a dynamical reconstruction of the dark energy equation of state
parameter by assuming that it satisfies a law of motion described by an
autonomous second-order differential equation, with the limit of the
cosmological constant as an equilibrium point. We determine the asymptotic
solutions of this equation and use them to construct two families of
parametric dark energy models, employing both linear and logarithmic
parametrization with respect to the scale factor. We perform observational
constraints by using the Supernova, the Cosmic Chronometers and the Baryon
Acoustic Oscillations of DESI DR2. The constraint parameters are directly
related with the initial value problem for the law of motion and its algebraic
properties. The analysis shows that most of the models fit the observational
data well with a preference to the models of the logarithmic parametrization.
Furthermore, we introduce a new class of models as generalizations of the CPL
model, for which the equilibrium point is a constant value rather than the
cosmological constant. These models fit the data in a similar or better way to
the CPL and the $\Lambda$CDM cosmological models.

\end{abstract}
\maketitle

\section{Introduction}

Recent cosmological data indicate that currently the universe is under an
acceleration phase \cite{rr1,Teg,Kowal,Komatsu,suzuki11}. Within the framework
of General Relativity, this late-fame expansion is attributed to the dark
energy fluid \cite{jo,jo1,jo2}, with a negative pressure component which
provides repulsive-gravitational forces within the universe. Nevertheless, the
lack of a fundamental physical theory explaining the mechanism behind cosmic
acceleration has led to the dark energy problem, for which a wide range of
cosmological scenarios and theories have been proposed for the description of
the cosmic acceleration.

The introduction of the cosmological constant $\Lambda$ within the
Einstein-Hilbert action is the simplest solution for the dark energy problem.
The cosmological constant, leading to the $\Lambda$CDM universe, has the
minimum number of free degrees of freedom and it can explain the late-time
acceleration phase. However, it suffers from mayor problems such as the fine
tuning problem and the coincidence problem \cite{Weinberg89,Pad03,Peri08}.
Recently, $\Lambda$CDM is challenged more due to the cosmological tensions
\cite{ten1,ten2,ten3}. Because of these problems, cosmologists have introduced
a plethora of theoretical
\cite{th0,th1,th2,th3,th4,th5,th6,th7,th7a,th7b,th7c,th7d,th7e,th7f,th7g} and
phenomenological proposals \cite{th8,th9,th10,th11,th12,th14} to address the
dark energy problem.

The determination of the equation of state parameter for the dark energy
component can provide crucial insights into the nature of this exotic fluid
source. Dynamical dark energy models with a time-varying equation of state
parameter $w\left(  a\right)  $ have been introduced to explain cosmological
observations. The Chevallier, Polarski and Linder \cite{cpl1,cpl1a} (CPL)
parametrization is the simplest varying dark energy model, introducing two
degrees of freedom. The CPL model helps to determine how distinct a given dark
energy model is from the cosmological constant \cite{cpl2,cpl3,cpl4,clp5}. Due
to the limitations of the CPL model, various alternative dark energy models
have been proposed over the past two decades. For more details, we refer the
reader to
\cite{par1,par2,par4,par5,par6,par7,exp1,exp2,osc1,osc2,osc3,osc4,osc5,pp1}
and the references therein. The recent Baryon Acoustic Oscillations of DESI
DR2 \ data indicate a preference for dynamical dark energy models over the
$\Lambda$CDM model. The family of $w_{0}w_{a}$CDM \cite{des4,des5,des6}. This
data set has been used to examined various theoretical models which lead to a
dynamical dark energy
\cite{bb1,bb2,bb3,bb4,bb5,bb6,bb7,bb8,bb9,bb10,bb11,bb12,bb13,bb14}.

In this study, we focus upon a dynamical reconstruction of the dark energy
equation of state parameter. Specifically, we assume that the equation of
state parameter satisfies a law of motion, meaning that it is governed by a
differential equation, with the requirement that the cosmological constant
limit exists as an equilibrium point at the present time. We determine all
possible functional forms for the asymptotic solution of this dynamical dark
energy law of motion near the equilibrium point, allowing us to construct
families of parametric dark energy models.

The parametrization of this law of motion is significant, as it provides
information about the nature of dark energy. We employ two types of
parametrization. Firstly, we consider a parametrization that is linear in the
scale factor, leading to exponential forms of the dark energy equation of
state parameter, with or without oscillatory components. These models exhibit
an approximate behavior similar to that of the CPL model. As a second
parametrization, we assume that the independent variable of this law of motion
follows a logarithmic function. This parametrization has a theoretical
foundation, as it can relate the evolution of dark energy parameters to the
dynamics of scalar fields \cite{sf1,sf2,sf3}, modified theories of gravity
\cite{sf4,sf4a} or interacting models \cite{sf5} and others. The dynamical
dark energy models related to the this parametrization have polynomial expression.

By using the above parametrization, we constructed two families of models and
applied observational constraints for the late universe to determine the free
parameters and assess the validity of the reconstruction approach. From the
data analysis, we found that most models fit the cosmological data well, in a
manner similar to the $\Lambda$CDM model. However, due to the larger number of
free parameters, $\Lambda$CDM remains statistically preferred. Nonetheless,
two models constructed using the logarithmic parametrization were found to be
statistically indistinguishable from $\Lambda$CDM. Furthermore, we found that
models with oscillatory components and decaying behavior are not favored by
the data.

Finally, we introduced a new class of models that can be seen as
generalizations of the CPL model, where the equilibrium point of the
differential equation is not the cosmological constant but a constant
parameter, introducing an additional degree of freedom within the models.
These new models fit the data as well as or better than the CPL model and are
statistically indistinguishable from it. This analysis provides important
insights into the law of motion that dark energy satisfies and its fundamental
nature. The structure of the paper is as follows.

In Section \ref{sec2} we introduce the cosmological model of our
consideration, which is that of spatially flat FLRW geometry in which the dark
energy has a dynamical equation of state parameter $w\left(  a\right)  $. In
Section \ref{sec2a} we assume that $w\left(  a\right)  $ is given by a
nonlinear differential equation $F\left(  w_{d},w_{d}^{\prime},w_{d}%
^{\prime\prime}\right)  =0$ in which the limit of the cosmological constant
$\Lambda$ exists as an equilibrium point at the present time. We employ the
theory of dynamical systems and explore the functional forms of $w\left(
a\right)  $ as it reaches the equilibrium point with the value $-1$.

In particular, we consider that $w\left(  a\right)  $ follows from a
second-order differential equation~$F\left(  w_{d},w_{d}^{\prime}%
,w_{d}^{\prime\prime}\right)  =0$ and we write three generic functional forms
of $w\left(  a\right)  $ related to the algebraic properties of the
differential equation $F\left(  w_{d},w_{d}^{\prime} ,w_{d}^{\prime\prime
}\right)  =0.$ These models depend upon four free parameters, except for the
case for which the differential equation $F$ is of first order and there are
two parameters. The determination of the free parameters of the models leads
to important information and constraints for the original physical law which
describes the dark energy equation of state parameter. However, from the
structure of the parameter space for the free variables, we discuss the
possible correlation of the free parameters. Therefore, we end with eight dark
energy models that depend upon one or two free parameters.

In Section \ref{sec3} we introduce the observational data applied in this
work. In particular, we consider the Pantheon+ Supernova data, the Cosmic
Chronometers and the Baryonic Acoustic Oscillators. The results from the
observational constraints of the proposed dynamical dark energy models are
presented in Section \ref{sec4}. Finally, in Section \ref{sec6}, we discuss
possible extension of this approach and we draw our conclusions.

\section{Cosmological Framework}

\label{sec2}

We assume that the physical space is described by a four-dimensional
Riemannian manifold with a metric tensor that is both isotropic and
homogeneous, as described by the spatially flat FLRW line element%
\begin{equation}
ds^{2}=-dt^{2}+a^{2}\left(  t\right)  \left(  dx^{2}+dy^{2}+dz^{2}\right)  ,
\label{cc.01}%
\end{equation}
where $a\left(  t\right)  $ is the scale factor which describes the radius of
the three-dimensional hypersurface.

Let $u^{\mu}=\delta_{t}^{\mu}$, with $u^{\mu}u_{\mu}=-1$,~be the comoving
observer. Let $T_{\mu\nu}$ be the energy momentum tensor for the cosmic fluid.
Within the $1+3$ decomposition we can define the observable energy density,
the pressure, the heat flux and the stress tensor as follows
\begin{equation}
\rho=T_{\mu\nu}u^{\mu}u^{\nu}, \label{cc.02}%
\end{equation}%
\begin{equation}
p=\frac{1}{3}T_{\mu\nu}h^{\mu\nu}, \label{cc.03}%
\end{equation}%
\begin{equation}
q_{\mu}=T_{\lambda\nu}h^{\lambda\nu}u_{\mu}, \label{cc.03b}%
\end{equation}%
\begin{equation}
\pi_{\mu\nu}=T_{\lambda\sigma}h_{\mu}^{\lambda}h_{\sigma}^{\lambda}-ph_{\mu
\nu}. \label{cc.03d}%
\end{equation}
in which $h_{\mu\nu}=g_{\mu\nu}+u_{\mu}u_{\nu}$ is the projective tensor.
Hence, the energy momentum tensor in terms of these latter physical quantities
is
\begin{equation}
T_{\mu\nu}=\rho u_{\mu}u_{\nu}+ph_{\mu\nu}+2q_{(\mu}u_{\nu)}+\pi_{\mu\nu}.
\label{cc.04}%
\end{equation}

For the gravitational model, we consider Einstein's General Relativity. The
gravitational field equations are given by
\begin{equation}
R_{\mu\nu}-\frac{1}{2}Rg_{\mu\nu}=T_{\mu\nu}, \label{cc.05}%
\end{equation}
where, for the line element (\ref{cc.01}) and the comoving observer, it
follows that%
\begin{align}
3H^{2}  &  =\rho,\label{cc.06}\\
2\dot{H}+3H^{2}  &  =p \label{cc.07}%
\end{align}
and $q_{\mu}=0$,$~\pi_{\mu\nu}=0$, where $H=\frac{\dot{a}}{a}$ is the Hubble
function and it is related to the expansion rate $\theta=u_{~;\mu}^{\mu}$ as
follows $\theta=3H$.

Within the FLRW geometry, the cosmic fluid is described by a perfect fluid.
Moreover, the Bianchi identity leads to the conservation law for the fluid
source:
\begin{equation}
\dot{\rho}+3H\left(  \rho+p\right)  =0. \label{cc.08}%
\end{equation}

We assume that the cosmological fluid consists of the following components:
\begin{equation}
\rho=\rho_{r}+\rho_{b}+\rho_{dm}+\rho_{d}, \label{cc.09}%
\end{equation}
where $\rho_{r}$ notes for the radiation, $p_{r}=\frac{1}{3}\rho_{r}$,
$\rho_{b}$ is the pressureless baryonic matter $p_{b}=0$, $\rho_{m}$ is the
cold dark matter, $p_{m}=0$, and $\rho_{d}$ marks as the energy density for
the dark energy source, $p_{d}=w_{d}\rho_{d}$. Therefore, the pressure
component of the cosmological fluid
\begin{equation}
p=\frac{1}{3}\rho_{r}+w_{d}\rho_{d}. \label{cc.10}%
\end{equation}

We substitute (\ref{cc.09}) and (\ref{cc.10}) into the continuity equation
(\ref{cc.08}). It follows that%
\begin{equation}
\left(  \dot{\rho}_{r}+4H\rho_{r}\right)  +\dot{\rho}_{b}+\dot{\rho}%
_{dm}+\left(  \dot{\rho}_{d}+3\left(  1+w_{d}\right)  H\rho_{d}\right)  =0.
\label{cc.11}%
\end{equation}
Furthermore, if there is no energy transfer between the four fluids, it
follows that
\begin{equation}
\rho_{r}=\rho_{r,0}\left(  \frac{a}{a_{0}}\right)  ^{-4},~\rho_{b}=\rho
_{b,0}\left(  \frac{a}{a_{0}}\right)  ^{-3},~\rho_{dm}=\rho_{dm,0}\left(
\frac{a}{a_{0}}\right)  ^{-3} \label{cc.12}%
\end{equation}
and%
\begin{equation}
\rho_{d}=\rho_{d,0}\exp\left(  -3\int_{a}^{a_{0}}\frac{1+w_{d}\left(
\alpha\right)  }{\alpha}d\alpha\right)  . \label{cc.13}%
\end{equation}
Parameters $\rho_{I,0}~$are integration constants and define the energy
density of the corresponding fluids at the present $\frac{a}{a_{0}}=1.$

With the use of these expressions, from equation (\ref{cc.06}) it follows that%
\begin{equation}
H\left(  a\right)  =H_{0}\sqrt{\Omega_{r0}\left(  \frac{a}{a_{0}}\right)
^{-3}+\Omega_{b0}\left(  \frac{a}{a_{0}}\right)  ^{-3}+\Omega_{dm0}\left(
\frac{a}{a_{0}}\right)  ^{-3}+\Omega_{d0}\exp\left(  -3\int_{a_{0}}^{a}%
\frac{1+w_{d}\left(  \alpha\right)  }{\alpha}d\alpha\right)  }, \label{cc.14}%
\end{equation}
in which $H_{0}$ is the value of the Hubble function at the present and
$\Omega_{I,0}=\frac{\rho_{I,0}}{3H_{0}^{2}}$.

From expression (\ref{cc.14}) we determine the algebraic equation%
\begin{equation}
\Omega_{r0}+\Omega_{b0}+\Omega_{dm0}+\Omega_{b0}=1. \label{cc.15}%
\end{equation}

The choice of the dark energy model determines the cosmic evolution of the
physical parameters. Simultaneously, the selection of the function
$w_{d}\left(  a\right)  $ is crucial for defining the cosmological model.

When the dark energy is described by the cosmological constant, it follows
that $w_{d}\left(  a\right)  =-1$, from which the $\Lambda$CDM model is
recovered. Expression (\ref{cc.14}) is%
\begin{equation}
H\left(  a\right)  =H_{0}\sqrt{\Omega_{r0}\left(  \frac{a}{a_{0}}\right)
^{-3}+\Omega_{b0}\left(  \frac{a}{a_{0}}\right)  ^{-3}+\Omega_{dm0}\left(
\frac{a}{a_{0}}\right)  ^{-3}+\Omega_{\Lambda0}}. \label{cc.16}%
\end{equation}
On the other hand, for a dark energy fluid with constant equation of state
parameter $w_{d}\left(  a\right)  =w_{0}$, the Hubble function is%
\begin{equation}
H\left(  a\right)  =H_{0}\sqrt{\Omega_{r0}\left(  \frac{a}{a_{0}}\right)
^{-3}+\Omega_{b0}\left(  \frac{a}{a_{0}}\right)  ^{-3}+\Omega_{dm0}\left(
\frac{a}{a_{0}}\right)  ^{-3}+\Omega_{d0}a^{-3\left(  1+w_{0}\right)  }}.
\label{cc.17}%
\end{equation}

The definition of the equation of state parameter, $w_{d}\left(  a\right)  , $
for the dark energy can follow from a theoretical background or a
phenomenological approach. Scalar fields or modified theories of gravity have
been introduced to explain the nature of dark energy and to construct dark
energy fluids that can account for cosmological observations. On the other
hand, within the phenomenological framework, parametric dark energy models
with specific functions, $w_{d}\left(  a\right)  , $ have been introduced in
the literature as an attempt to understand the effects of $w_{d}\left(
a\right)  $ in the analysis of cosmological parameters.

The Chevallier-Polarski-Linder (CPL) model is the simplest parametrization of
the $w_{d}\left(  a\right)  $, where $w_{d}\left(  a\right)  $ is a linear
function such that $w_{d}\left(  a\right)  =w_{0}+w_{a}\left(  1-\left(
\frac{a}{a_{0}}\right)  \right)  $ or, equivalently,
\begin{equation}
w_{d}^{CPL}\left(  z\right)  =w_{0}+w_{a}\left(  \frac{z}{1+z}\right)  ,
\end{equation}
where $1+z=\frac{1}{a}$ is the redshift parameter. The CPL parametrization
provides the two limits. At the present $w_{d}\left(  z\rightarrow0\right)
=w_{0}$, while in the past $w_{d}\left(  z\rightarrow+\infty\right)
=w_{0}+w_{a}$.

The CPL model can be seen as the fist-order approximation of the exponential
parametric dark energy model $w_{d}\left(  a\right)  =w_{0}+w_{a}\left(
e^{1-a}-1\right)  $~or, equivalently~\cite{exp1},%
\begin{equation}
w_{d}^{Exp}\left(  z\right)  =w_{0}+w_{a}e^{\frac{z}{1+z}}\text{.}%
\end{equation}
Indeed, if we expand the latter in terms of the Taylor expansion, it follows
that~\cite{exp2}%
\begin{equation}
w_{d}^{Exp}\left(  z\right)  =w_{0}+w_{a}\left(
%TCIMACRO{\dsum \limits_{n=1}^{\infty}}%
%BeginExpansion
{\displaystyle\sum\limits_{n=1}^{\infty}}
%EndExpansion
\left[  \frac{1}{n!}\left(  \frac{z}{1+z}\right)  ^{n}\right]  \right)  ,
\end{equation}
where in the first order approximation, for $z<<1$, it follows that~$w_{d}%
^{Exp}\left(  z\right)  \simeq w_{d}^{CPL}\left(  z\right)  $.

Dark energy models with oscillating terms \cite{osc1,osc2,osc3,osc4,osc5} have
been proposed in the literature, see, for instance, the model \cite{osc5}%
\begin{equation}
w_{d}^{Osc}\left(  z\right)  =-1+w_{a}\frac{z}{1+z^{2}}\sin\left(
w_{b}z\right)  ,
\end{equation}
which has the $\Lambda$ limit at the present $w_{d}^{Osc}\left(
z\rightarrow0\right)  =-1$ and in the past, $w_{d}^{Osc}\left(  z\rightarrow
+\infty\right)  =-1$. In the first order approximation for low values of the
redshift it follows that $w_{d}^{Osc}\left(  z\right)  \simeq w_{d}%
^{CPL}\left(  z,w_{0}\rightarrow-1\right)  $, which is that of the CPL model
with $w_{0}=-1$.

\section{Dynamical dark energy with $\Lambda$ as an Equilibrium Point}

\label{sec2a} There is a \textquotedblleft zoology\textquotedblright\ of
proposed models in the literature as we discussed in the Introduction.
\ However, in this work, we consider that the dark energy equation of state
parameter $w_{d}\left(  a\right)  $ satisfies an autonomous differential
equation of the form $F\left(  w_{d},w_{d}^{\prime},w_{d}^{\prime\prime
},...\right)  =0$, where $w_{d}^{\prime}$ denotes the derivative with respect
to the scale factor, such that $w_{d}^{\prime}=\frac{dw_{d}\left(  a\right)
}{d\chi\left(  a\right)  }$, where $\chi\left(  a\right)  $ is a smooth,
continuous function. This approach is motivated by certain scalar field models
that have been used to describe inflation; see, for instance, \cite{jf2,jf3}.

Physics laws are often described by second-order differential equations.
Inspired by this, we assume that the differential equation $F\left(
w_{d},w_{d}^{\prime},w_{d}^{\prime\prime}\right)  =0$ is of second order and
takes the form
\begin{equation}
w_{d}^{\prime\prime}=W\left(  w_{d},w_{d}^{\prime}\right)  , \label{cc.18}%
\end{equation}
where $w_{d}=-1$ is an equilibrium point, meaning that $W\left(  w_{d}%
,w_{d}^{\prime}\right)  {|w{d}\rightarrow-1}=0$, ensuring that the equilibrium
constant is well-defined.

In this work we do not define explicitly the differential equation that the
equation of state parameter satisfies. But we try to reconstruct the solution
of the equation of state parameter near to the equilibrium point. For
second-order theories, like quintessence, or phantom scalar field theories,
the equation that $w_{d}\left(  \chi\left(  a\right)  \right)  $ satisfies is
a first-order differential equation. On the other hand, for higher-order
theories of gravity the order of differential equation is higher.

We write equation (\ref{cc.18}) in the equivalent form%
\begin{equation}
w_{d}^{\prime}=P_{w},~~P_{w}^{\prime}=W\left(  w_{d},P_{w}\right)
\label{cc.19}%
\end{equation}
and assume that the equilibrium $Q=\left(  w_{d}\left(  Q\right)
,P_{w}\left(  Q\right)  \right)  $ has components $Q=\left(  -1,P_{w}%
^{0}\right)  $. Then around the stationary point, the asymptotic solution for
the equation of state parameter is%
\begin{equation}
w_{d}\left(  a\right)  =-1+w_{d}^{1}e^{-\Delta_{+}\left(  \chi\left(
a\right)  -\chi\left(  a_{0}\right)  \right)  }+w_{d}^{2}e^{-\Delta_{-}\left(
\chi\left(  a\right)  -\chi\left(  a_{0}\right)  \right)  }+c, \label{cc.20}%
\end{equation}
in which $w_{d}^{1}$ and $w_{d}^{2}$ are two constants of integration and%
\begin{equation}
\Delta_{\pm}=\frac{1}{2}\left[  \left(  \frac{\partial W}{\partial P_{w}%
}\right)  \pm\sqrt{\left(  \frac{\partial W}{\partial P_{w}}\right)
^{2}+4\left(  \frac{\partial W}{\partial w_{d}}\right)  }\right]  _{\left(
w_{d},P_{w}\right)  \rightarrow\left(  -1,P_{w}^{0}\right)  }. \label{cc.21}%
\end{equation}
Constant $c$ has been added, such that at the equilibrium point at the present
time $w_{d}\left(  a_{0}\right)  =-1$. Its value depends upon the function
$\chi\left(  a\right)  $. Indeed, for $\chi=\ln\left(  \frac{a}{a_{0}}\right)
$, it is easy to see that $c=0$.

We consider the following case three cases for the value of the determinant at
the equilibrium point, (i) $\left[  \left(  \frac{\partial W}{\partial P_{w}%
}\right)  ^{2}+4\left(  \frac{\partial W}{\partial w_{d}}\right)  \right]
=0$, (ii)$~\left(  \frac{\partial W}{\partial P_{w}}\right)  ^{2}+4\left(
\frac{\partial W}{\partial w_{d}}\right)  >0$ and (iii) $\left(
\frac{\partial W}{\partial P_{w}}\right)  ^{2}+4\left(  \frac{\partial
W}{\partial w_{d}}\right)  <0$.

When the two roots $\Delta_{+}$ and $\Delta_{-}$ are equal, the asymptotic
solution around the equilibrium point is%
\begin{equation}
w_{d}\left(  a\right)  =-1+w_{d}^{1}e^{-w_{a}\left(  \chi\left(  a\right)
-\chi\left(  a_{0}\right)  \right)  }~,~ \label{cc.22}%
\end{equation}
where $w_{a}=\Delta_{+}=\Delta_{-}$.

For the second case, in which the two roots are real, it follows that the
asymptotic solution around the equilibrium point $Q$ is%
\begin{equation}
w_{d}\left(  a\right)  =-1+w_{d}^{1}\exp\left[  -w_{a}\left(  \chi\left(
a\right)  -\chi\left(  a_{0}\right)  \right)  \right]  +w_{d}^{2}\exp\left[
-w_{b}\left(  \chi\left(  a\right)  -\chi\left(  a_{0}\right)  \right)
\right]  +c, \label{cc.23}%
\end{equation}
with $w_{a}=\Delta_{+}$ and $w_{b}=\Delta_{-}$.

Finally, when the two eigenvalues are complex, the asymptotic solution is
expressed as%
\begin{align}
w_{d}\left(  a\right)   &  =-1+w_{d}^{1}\exp\left(  -w_{a}\left(  \chi\left(
a\right)  -\chi\left(  a_{0}\right)  \right)  \right)  \sin\left[
w_{b}\left(  \chi\left(  a\right)  -\chi\left(  a_{0}\right)  \right)  \right]
\nonumber\\
&  +w_{d}^{2}\exp\left[  -\left(  w_{a}\left(  \chi\left(  a\right)
-\chi\left(  a_{0}\right)  \right)  \right)  \right]  \cos\left[  w_{b}\left(
\chi\left(  a\right)  -\chi\left(  a_{0}\right)  \right)  \right]  +c,
\label{cc.24}%
\end{align}
where now $w_{a}=\operatorname{Re}\left(  \Delta_{+}\right)
=\operatorname{Re}\left(  \Delta_{-}\right)  $ and $w_{b}=\operatorname{Im}%
\left(  \Delta_{+}\right)  =\operatorname{Im}\left(  \Delta_{-}\right)  $. For
a geometric interpretation of the stationary points we refer the reader in
\cite{stro1}.

The requirement that the $\Lambda$CDM model be a future attractor implies that
the parameter $w_{a}$ must always be positive, i.e., $w_{a}>0$. In the second
case, the additional condition $w_{b}>0$ must also hold. On the other hand,
when the real parts of the eigenvalues are positive, the $\Lambda$CDM model
behaves as a source. When the real parts of both eigenvalues are zero, the
equilibrium point is a center, and the asymptotic solution for the equation of
state parameter oscillates around the value $-1$.

These solutions describe the asymptotic behavior near equilibrium points.
However, in the following, we use them as motivation to define parametric dark
energy models.

At this point, it is important to note that, if the parametric dark energy
equation of state parameter $w_{d}\left(  a\right)  $ were described by a
first-order differential equation $F\left(  w_{d},w_{d}^{\prime}\right)  =0$,
then only the asymptotic behavior given by (\ref{cc.22}) could be constructed.
In contrast, for higher-order differential equations, the number of
eigenvalues increases, thereby introducing additional degrees of freedom.

Dynamical dark energy models with the above asymptotic behaviours can be
easily reconstructed in single- and multi- scalar field cosmological models
\cite{sf1,sf2,sf3}.

\subsection{Linear function $\chi\left(  a\right)  =\frac{a}{a_{0}}$.}

Consider now the linear function $\chi\left(  a\right)  =\frac{a}{a_{0}}$,
where $\chi\left(  1\right)  =1$. \ We define the three parametric dark energy
models
\begin{equation}
w^{I}\left(  z\right)  =-1+w_{d}^{1}\left(  \exp\left(  w_{a}Z\right)
-1\right)  , \label{cc.25}%
\end{equation}%
\begin{equation}
w^{II}\left(  z\right)  =-1+w_{d}^{1}\left(  \exp\left(  w_{a}Z\right)
-1\right)  +w_{d}^{2}\left(  \exp\left(  w_{b}Z\right)  -1\right)  ,
\label{cc.26}%
\end{equation}
and\qquad%
\begin{equation}
w^{III}\left(  z\right)  =-1+w_{d}^{1}\left[  \exp\left(  w_{a}Z\right)
\sin\left(  w_{b}Z+w_{c}\right)  -\sin\left(  w_{c}\right)  \right]  .
\label{cc.27}%
\end{equation}
with $Z=\frac{z}{1+z}$ and parameter $c$ has been defined. Parameter $Z$ takes
values within the range $[0,1)$, where $0$ corresponds to the present value
and $1$ to the limit $a\rightarrow0$. \ 

At the present, i.e. $Z<<0$, the three models takes the form of the CPL model
for $w_{0}=-1$, that is,%
\begin{align}
w^{I}\left(  z\rightarrow0^{+}\right)   &  \simeq-1+\left(  w_{d}^{1}%
w_{a}\right)  Z,\\
w^{II}\left(  z\rightarrow0^{+}\right)   &  \simeq-1+\left(  w_{d}^{1}%
w_{a}+w_{d}^{2}w_{b}\right)  Z,\\
w^{III}\left(  z\rightarrow0^{+}\right)   &  \simeq-1+\left(  w_{d}^{1}\left(
w_{b}\sin\left(  w_{c}\right)  -\sin\left(  w_{c}\right)  \right)  \right)  Z.
\end{align}

In the limit where $Z\rightarrow1$, the equation of state parameters gives the
values%
\begin{align}
w^{I}\left(  z\rightarrow+\infty\right)   &  \simeq-1+w_{d}^{1}\left(
\exp\left(  w_{a}\right)  -1\right)  ,\\
w^{II}\left(  z\rightarrow+\infty\right)   &  \simeq-1+w_{d}^{1}\left(
\exp\left(  w_{a}\right)  -1\right)  +w_{d}^{2}\left(  \exp\left(
w_{b}\right)  -1\right)  ,\\
w^{III}\left(  z\rightarrow+\infty\right)   &  \simeq-1+w_{d}^{1}\left[
\sin\left(  w_{b}+w_{c}\right)  -\sin\left(  w_{c}\right)  \right]  .
\end{align}
We remark that for $w_{a}=1$, function $w^{I}\left(  z\right)  $ recovers the
exponential parametric dark energy model studied in \cite{exp2}.

\subsection{Logarithmic function $\chi\left(  a\right)  =\ln\left(  \frac
{a}{a_{0}}\right)  $}

A second set of common parametrization is the logarithmic function
$\chi\left(  a\right)  =\ln\left(  \frac{a}{a_{0}}\right)  $. This choice
follows naturally within the Hubble normalization \cite{sf1}, which is
frequently applied in the study of the dynamics of gravitational field
equations \cite{sf1,sf2,sf3,sf4,sf5}.

Within the logarithmic parametrization we define the parametric dark energy
models%
\begin{equation}
w^{IV}\left(  z\right)  =-1+w_{d}^{1}~Z^{w_{a}}, \label{cc.28}%
\end{equation}%
\begin{equation}
w^{V}\left(  z\right)  =-1+w_{d}^{1}~Z^{w_{a}}+w_{d}^{2}~Z^{w_{b}},
\label{cc.29}%
\end{equation}
and%
\begin{equation}
w^{VI}\left(  z\right)  =-1+w_{d}^{1}\left[  Z^{w_{a}}\sin\left(  w_{b}%
Z+w_{c}\right)  -\sin\left(  w_{c}\right)  \right]  . \label{cc.30}%
\end{equation}
In this formulation, in order the models to be well defined, it is necessary
that $w_{a},~w_{b}>0$, while the CPL limit is recovered only in the models
$w^{IV}$ and $w^{VI}$ for $w_{a}=1$.

On the other hand, for $Z=1$, the parametric functions $w^{IV}\left(
z\rightarrow+\infty\right)  $ and~$w^{V}\left(  z\rightarrow+\infty\right)  $
have the limit $-1$, while function $w^{VI}$ gives the value%
\begin{equation}
w^{VI}\left(  z\rightarrow+\infty\right)  =-1+w_{d}^{1}\left(  \sin\left(
w_{b}+w_{c}\right)  -\sin\left(  w_{c}\right)  \right)  .
\end{equation}

The parametric model $w^{IV}\left(  z\right)  $ was previously introduced in
\cite{pant1} and is known as the nCPL model, where $n$ refers to the power of
$w_{a}$. Observational analysis using the Union 2.1 dataset indicated a
preference for high values of the parameter $n$. In this work, we will
investigate whether this preference persists with more recent cosmological data.

We note that the parameters $w_{a}$ and $w_{b}$ explicitly depend on the
underlying theoretical framework, specifically on the differential equations
that govern $w_{d}\left(  a\right)  $. However, the parameters $w_{d}^{1}$ and
$w_{d}^{2}$ depend solely on the initial conditions.

\subsection{Parametric space}

\begin{figure}[ptbh]
\centering\includegraphics[width=1\textwidth]{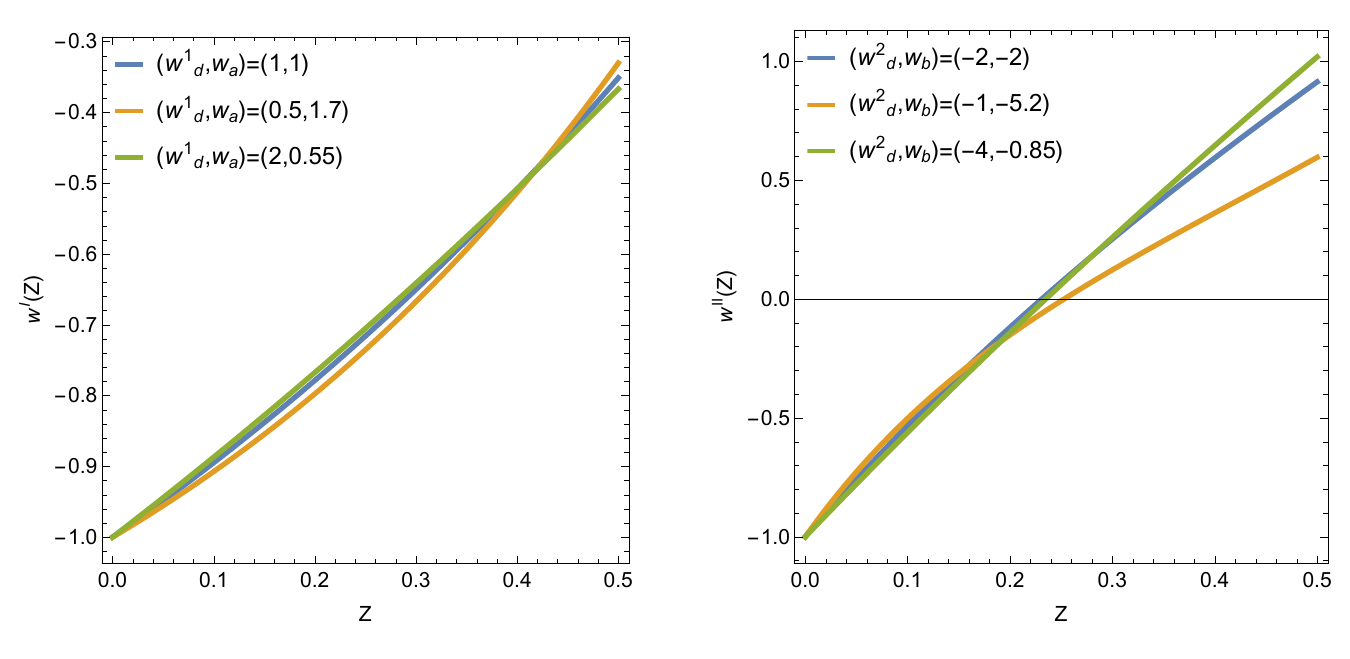}\caption{Qualitative
evolution of $w^{I}\left(  z;w_{d}^{1},w_{a}\right)  ~$(Left Fig.) and
$w^{II}\left(  z;w_{d}^{1},w_{a},w_{d}^{2},w_{b}\right)  $ (Right Fig.) for
different values of the parameters. We observe that for small $Z$ the sets of
parameters $\left(  w_{d}^{1},w_{a}\right)  $ and $\left(  w_{d}^{2}%
,w_{b}\right)  $ are degenerate.}%
\label{plot1}%
\end{figure}

We discuss the parametric space for the free variables that we shall use
within the analysis for the comparison to the observation data. Parametric
model $w^{I}\left(  z\right)  $ depends on two parameters $w_{d}^{1}$ and
$w_{a}$, however, for small redshift we found that the two parameters are
degenerate. Indeed, in terms of statistics it will not be possible to constant
these two parameters. The same discussion hold and for models $w^{II}\left(
z\right)  $ and $w^{III}\left(  z\right)  $. In Fig. \ref{plot1} we present
the qualitative evolution of parametric models $w^{I}\left(  z\right)  $, and
$w^{II}\left(  z\right)  $ from where we observe that for small $z$,
parameters are related as $w_{a}\sim\frac{1}{w_{d}^{1}}$ and $w_{b}\sim
\frac{1}{w_{d}^{2}}$.

If in model $w^{I}\left(  z\right)  $ we assume $w_{a}=1$, then we recover the
model studied in \cite{exp2}. In contrary, in this study we will reduce the
parametric space by assuming $w_{d}^{1}=\pm1\,$~in$~w^{I}\left(  z\right)  $
and $w_{d}^{1}=-w_{d}^{2}=1$ in$~w^{II}\left(  z\right)  $; that is, we end
with the following dynamical dark energy equation of state parameters%
\begin{align*}
w_{+}^{I}\left(  z\right)   &  =-1+e^{w_{a}Z}-1,\\
w_{-}^{I}\left(  z\right)   &  =-1-\left(  e^{w_{a}Z}-1\right)  ,\\
w^{II}\left(  z\right)   &  =-1+\sinh\left(  w_{a}Z\right)  .
\end{align*}
In $w^{II}\left(  z\right)  $ we have assumed that $w_{d}^{1},~w_{d}^{2}$ have
different signs in order to avoid any degeneration on the indices.

For model $w^{III}\left(  z\right)  $, and parameter $w_{c}$ we consider the
extreme limit$~w_{c}=0$, thus
\begin{equation}
w^{III}\left(  z\right)  =-1+\exp\left(  w_{a}Z\right)  \sin\left(
w_{b}Z\right)  .
\end{equation}
This model allows can change sign.

We conclude that models $w^{I}\left(  z\right)  $,~$w^{II}\left(  z\right)  $
depend on one parameter, namely $w_{a}$, and model$~w^{III}\left(  z\right)  $
on two parameters, that is, $w_{a}$ and $w_{b}$. Although in the first-order
approximation of$~w^{III}\left(  z\right)  $, parameters $w_{a},~w_{b}$ are
independent, this is not true in the higher-order approximation. As we shall
see later, a correlation exists between these two parameters.

For the second family of parametric models $w^{IV}\left(  z\right)  $,
$w^{V}\left(  z\right)  $ parameters $\left(  w_{d}^{1},~w_{a}\right)  $ and
$\left(  w_{d}^{2},~w_{b}\right)  $ maybe are not related explicitly as before
for small values of redshifts $z$. Nevertheless, for small values of $z$, as
$w_{a}$ increases~$\uparrow$, the term~$Z^{w_{a}}~\downarrow$. Then, if
$w_{d}^{1}~\uparrow$, the product changes very slowly. These variables are
degenerate for small redshifts.

Therefore we introduce the following models%
\begin{align}
w_{+}^{IV}\left(  z\right)   &  =-1+Z^{w_{a}},\\
w_{-}^{IV}\left(  z\right)   &  =-1-Z^{w_{a}},\\
w^{V}\left(  z\right)   &  =-1+Z^{w_{a}}-~Z^{w_{b}},
\end{align}
where again for model $w^{V}\left(  z\right)  ~$we assume that the constants
of integration related to the initial conditions to have different signs.

Finally, from the generic function of $w^{VI}\left(  z\right)  ,$ similarly as
above we assume $w_{c}=0$. We end up with the parametric dark energy model
\begin{equation}
w_{0}^{VI}\left(  z\right)  =-1+Z^{w_{a}}\sin\left(  w_{b}Z\right)  .
\end{equation}

The dynamical dark energy models studied in this work are summarized in Table
\ref{tab1}.%

%TCIMACRO{\TeXButton{B}{\begin{table}[tbp] \centering}}%
%BeginExpansion
\begin{table}[tbp] \centering
%EndExpansion
\caption{Dynamical Dark Energy Models}%
\begin{tabular}
[c]{ccc}\hline\hline
\textbf{Model~} & \textbf{Parameters} & \textbf{Function }$\left(
\mathbf{Z=}\frac{\mathbf{z}}{\mathbf{1+z}}\right)  $\\\hline
$w_{+}^{I}\left(  z\right)  $ & $w_{a}$ & $-1+\left(  e^{w_{a}Z}-1\right)  $\\
$w_{-}^{I}\left(  z\right)  $ & $w_{a}$ & $-1-\left(  e^{w_{a}Z}-1\right)  $\\
$w^{II}\left(  z\right)  $ & $w_{a}$ & $-1+\sinh\left(  w_{a}Z\right)  $\\
$w^{III}\left(  z\right)  $ & $w_{a},w_{b}$ & $-1+\exp\left(  w_{a}Z\right)
\sin\left(  w_{b}Z\right)  $\\
$w_{+}^{IV}\left(  z\right)  $ & $w_{a}$ & $-1+~Z^{w_{a}}$\\
$w^{IV}\left(  z\right)  $ & $w_{a}$ & $-1-~Z^{w_{a}}$\\
$w^{V}\left(  z\right)  $ & $w_{a},w_{b}$ & $-1+Z^{w_{a}}-~Z^{w_{b}}$\\
$w^{VI}\left(  z\right)  $ & $w_{a},w_{b}$ & $-1+Z^{w_{a}}\sin\left(
w_{b}Z\right)  $\\\hline\hline
\end{tabular}
\label{tab1}%
%TCIMACRO{\TeXButton{E}{\end{table}}}%
%BeginExpansion
\end{table}%
%EndExpansion

\section{Observational Data}

\label{sec3}

For our analysis, in order to determine the cosmological parameters, we make
use of three background datasets: the Pantheon+ Type Ia supernova data (SN),
the Cosmic Chronometers (CC) and the Baryonic Acoustic Oscillations (BAO) data.

\begin{itemize}
\item (SN) The Pantheon+
dataset\footnote{https://github.com/PantheonPlusSH0ES/DataRelease} comprises
1701 light curves of 1550 spectroscopically confirmed supernova events within
the range $10^{-3}<z<2.27~$\cite{pant}. The data provides the distance modulus
$\mu^{obs}~$at~observed redshifts~$z$. The theoretical distance modulus is
constructed as follows~$\mu^{th}=5\log D_{L}+25,$ where $D_{L}$ is the
luminosity distance defined by the Hubble functions from the expression
$D_{L}=c\left(  1+z\right)  \int\frac{dz}{H\left(  z\right)  }$.

\item (CC) For the Cosmic Chronometers we use 31 direct measurements of the
Hubble parameter across redshifts in the
range\footnote{https://github.com/Ahmadmehrabi/Cosmic\_chronometer\_data}
$0.09\leq z\leq1.965~$ as summarized in Table 1 in \cite{sunny}.

\item (BAO) The BAO data are provided by the SDSS Galaxy Consensus, quasars
and Lyman-$\alpha$ forests \cite{bbn0} and the data from the DESI DR1
collaboration \cite{des1,des2,des3} where we refer as BAO$_{1}$ and the DESI
DR2 collaboration \cite{des4,des5,des6} which we refer as BAO$_{2}$. These
data sets provide observation values of the $\frac{D_{M}}{r_{s}},~\frac{D_{V}%
}{r_{s}}$, $\frac{D_{H}}{r_{s}}$ where $r_{s}$ is the sound horizon at the
drag epoch, $D_{M}$ is the comoving angular distance $D_{M}=\left(
1+z\right)  ^{-1}D_{L}$ ; $D_{V}$ is the volume averaged distance $D_{V}%
^{3}=cD_{L}\frac{z}{H\left(  z\right)  }$ and $D_{H}$ is the Hubble distance
$D_{H}=\frac{c}{H}$.
\end{itemize}

\subsection{Methodology}

In order to obtain the optimal values for the cosmological parameters we
employ the Bayesian inference COBAYA\footnote{https://cobaya.readthedocs.io/}
\cite{cob1,cob2} with a custom theory and the MCMC sampler. From the results
of the\ Planck 2018 collaborations \cite{pl1n} we consider the energy density
for the radiation to be $\Omega_{r0}=4.15~10^{-5}$. Thus, we constrain our
proposed models for the parameters $\left\{  H_{0},\Omega_{m0},\mathbf{w,}%
r_{drag}\right\}  ,~$where $\mathbf{w}$ describes the vector of parameters
within the equation of state parameters, as they are given in Table \ref{tab1}
and $\Omega_{m0}=\Omega_{dm0}+\Omega_{b0}$.

For all the runs we selected the priors $H_{0}\in\left[  60,75\right]
$,$~\Omega_{m0}\in\left[  0.2,0.45\right]  $ and $r_{drag}\in\left[
130,160\right]  $.

The parameters in $\mathbf{w}$ were assigned different ranges for each model,
because their contributions vary due to the different functional forms of the
dark energy models.

As the models under consideration have different dimension $\kappa$ in their
parameter space, we introduce the Akaike Information Criterion (AIC) to
compare them statistically. The AIC, for Gaussian errors, is defined as
\[
AIC=-2\ln\mathcal{L}_{\max}+2\kappa+\frac{2\kappa\left(  \kappa-1\right)
}{N_{tot}-\left(  \kappa+1\right)  },
\]
where~$\mathcal{L}_{\max}$ is the maximum value for the likelihood and
$N_{tot}$ is the total number of data.\ For a large number of data the later
formula is%
\[
AIC\simeq-2\ln\mathcal{L}_{\max}+2\kappa
\]

Recall that for the combined data, the likelihood $\mathcal{L}_{\max}$ is
defined as%
\[
\mathcal{L}_{\max}^{total}=\mathcal{L}_{\max}^{SN}\times\mathcal{L}_{\max
}^{CC}\times\mathcal{L}_{\max}^{BAO},
\]
where $\mathcal{L}_{\max}=\exp\left(  -\frac{1}{2}\chi_{\min}^{2}\right)  $.
The difference of the AIC parameters between two models%
\[
\Delta\left(  AIC\right)  =AIC_{1}-AIC_{2}%
\]
gives information if the models are statistically indistinguishable. The
higher the value of parameter $\Delta AIC$, the higher the evidence against
the model with higher value of $AIC$; a difference $\left\vert \Delta
AIC\right\vert >2~$ indicates a positive such evidence and $\left\vert \Delta
AIC\right\vert >6$ indicates a strong such evidence, while a value $\left\vert
\Delta AIC<2\right\vert ~$ indicates consistency among the two models in comparison.

\section{Results}

\label{sec4}

We constrain the dynamical dark energy models using two sets of data: (I)
Supernova and Cosmic Chronometers ($SN+CC$) and (II) Supernova, Cosmic
Chronometers and Baryonic Acoustic Oscillations of DESI DR1 ($SN+CC+BAO_{1}$)
and (III) Supernova, Cosmic Chronometers and Baryonic Acoustic Oscillations of
DESI DR2~($SN+CC+BAO_{2}$).~We determine the free parameters along with their
corresponding $1\sigma$ errors by locating the maximum of the likelihood
functions. The results for the two tests are presented in Tables \ref{tab2A},
\ref{tab2b} and \ref{tab2c}.

\begin{figure}[ptbh]
\centering\includegraphics[width=1\textwidth]{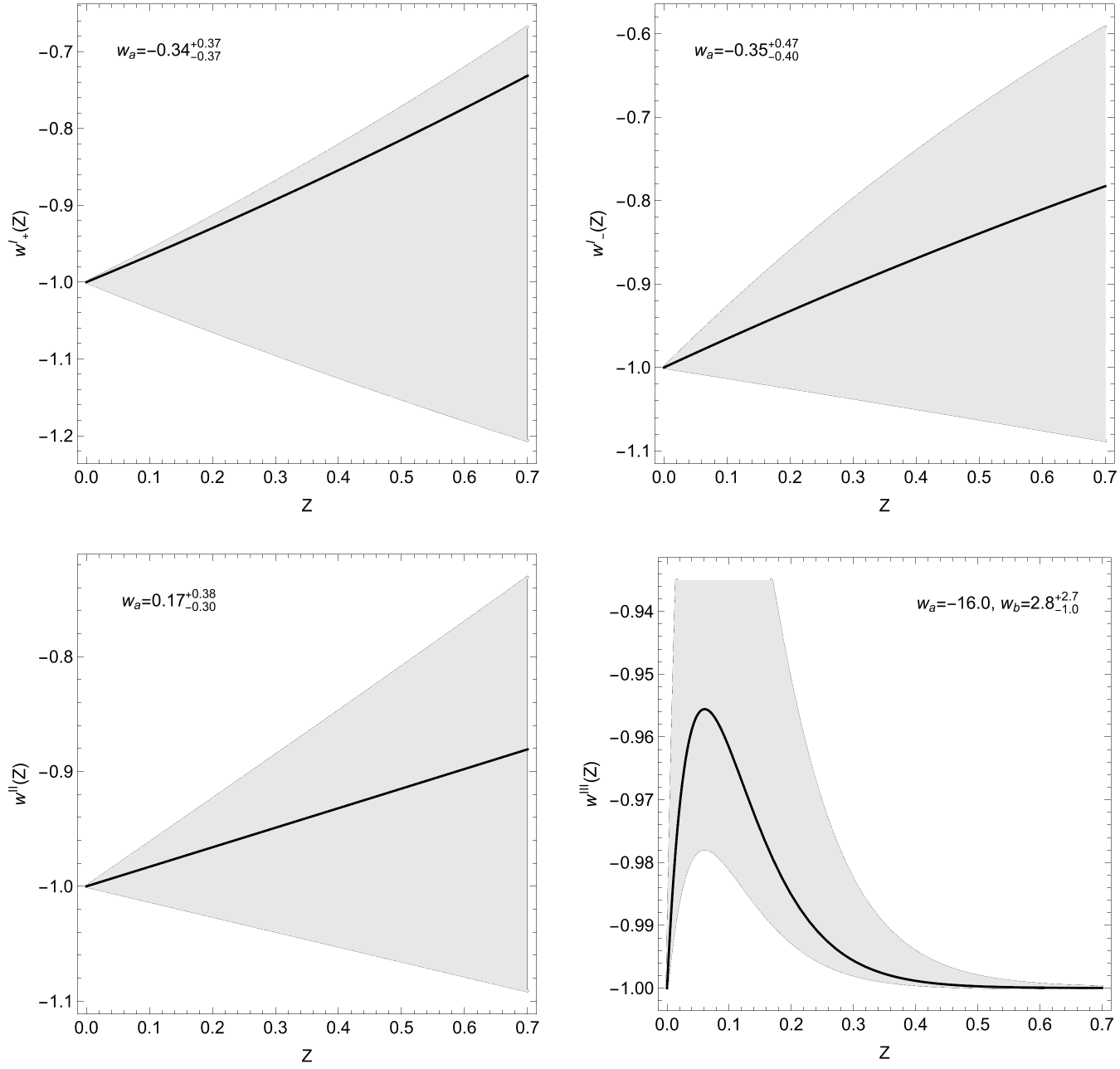}\caption{Evolution for
the dark energy equation of state parameter $w_{\pm}^{I}\left(  z\right)
,w^{II}\left(  z\right)  $ and $w^{III}\left(  z\right)  $.}%
\label{fig01}%
\end{figure}

\begin{figure}[ptbh]
\centering\includegraphics[width=1\textwidth]{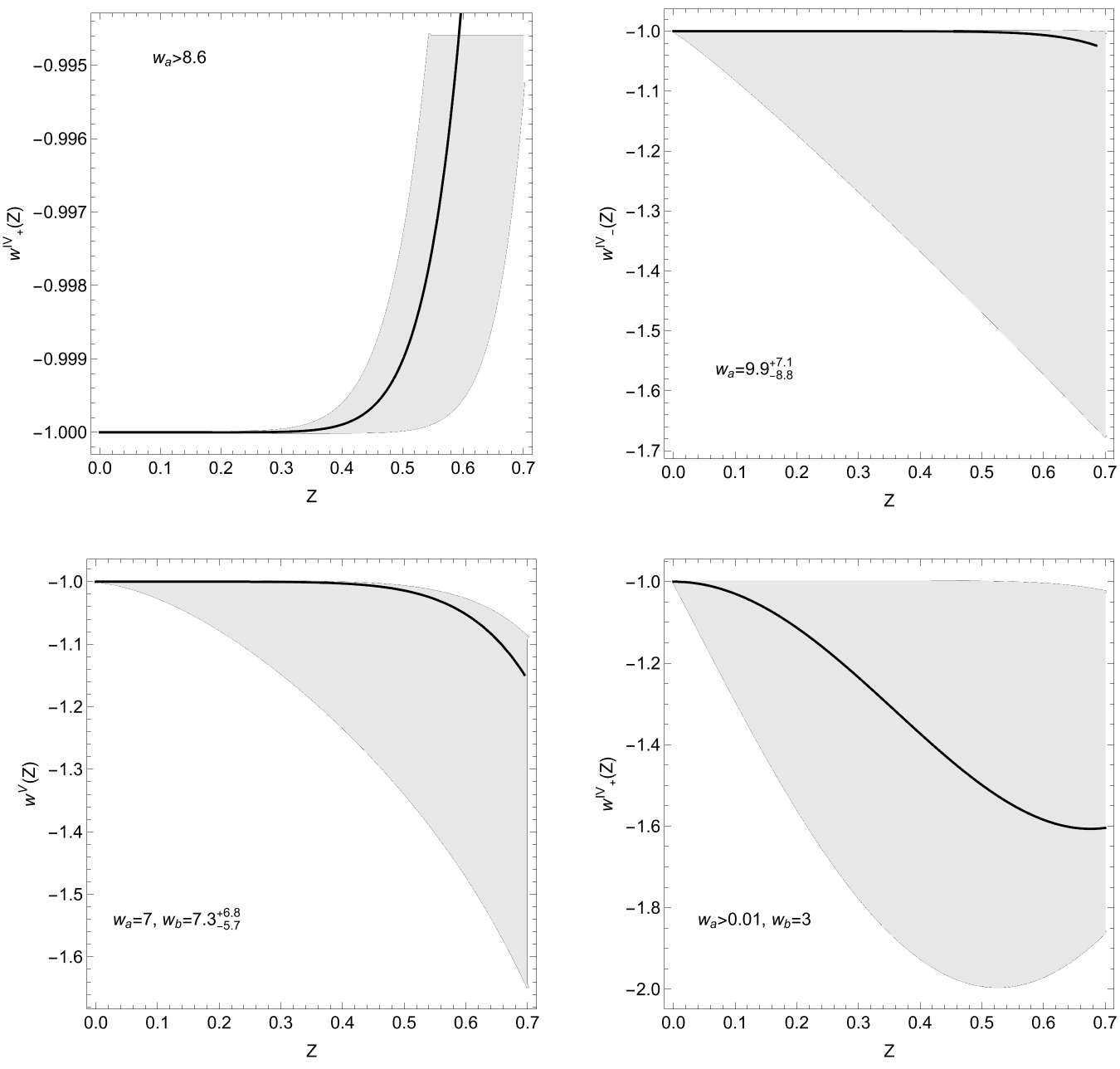}\caption{Evolution for
the dark energy equation of state parameter $w_{\pm}^{IV}\left(  z\right)
,w^{V}\left(  z\right)  $ and $w^{VI}\left(  z\right)  $.}%
\label{fig02}%
\end{figure}

We consider the $\Lambda$CDM model~as a reference theory. In Tables
\ref{tab2A}, \ref{tab2b} and \ref{tab2c}, we compute the parameter
$\Delta\left(  AIC\right)  $ for each constraint to assess the statistical
preference of the models based on the given datasets.~We note that the models
$w_{\pm}^{I}\left(  z\right)  ,~w^{II}\left(  z\right)  $ and $w_{\pm}%
^{IV}\left(  z\right)  $ introduce only one additional degree of freedom
compared to $\Lambda$CDM, whereas the models $w^{III}\left(  z\right)  $,
$w^{V}\left(  z\right)  $, and $w^{VI}\left(  z\right)  $ have two extra free
parameters. In Figs. \ref{fig01} and \ref{fig02} we present the qualitative
evolution of the dynamical dark energy parameters.

\subsection{$SN+CC$ data}

From the $SN+CC$ data, the $\Lambda$CDM model provides the maximum likelihood
with the minimum value $\chi_{SN+CC}^{2}\left(  \Lambda\right)  =1419.5$. The
models $w_{\pm}^{I}\left(  z\right)  $ fit the data with similar values for
the physical parameters, with best-fit values $w_{a}^{I-}=-0.24$ and
$w_{a}^{I+}=-0.05$, and a corresponding chi-square value of $\chi_{SN+CC}%
^{2}\left(  w_{+}^{I}\right)  =1419.8$ and $\chi_{SN+CC}^{2}\left(  w_{+}%
^{I}\right)  =1419.9$. Thus, the difference in the Akaike Information
Criterion (AIC) compared to $\Lambda$CDM are $\Delta\left(  AIC\right)  =2.3$
and $2.4$ respectively.

For model $w^{II}\left(  z\right)  $, the best-fit value is $w_{a}=-0.37$,
with a minimum chi-square value of $\chi_{SN+CC}^{2}\left(  w^{II}\right)
=1419.9$, leading to $\Delta\left(  AIC\right)  =2.4$.~On the other hand, for
model $w^{III}\left(  z\right)  $, we search for the best-fit values within
the ranges $w_{a}\in\left[  -30,10\right]  $ and $w_{b}\in\left[  -5,5\right]
$. The analysis indicates that the two variables are degenerate, as previously
discussed. Nevertheless, within this range, we find that $w_{a}<-12.6$ and
$w_{b}=1.1$, with $\chi_{SN+CC}^{2}\left(  w^{III}\right)  =1419.7$ and
$\Delta\left(  AIC\right)  =4.2$.

For the models of the second parametrization, we consider the parameter ranges
$w_{a}\in(0,30]$ and $w_{b}\in\lbrack0,30]$ or $w_{b}\in\lbrack-10,10]$.
However, this dataset is not sufficient to fully constrain the free parameters
of this family of models.~For models $w_{\pm}^{IV}\left(  z\right)  $, we find
that $w_{a}>6.6$ and $w_{a}=10.1$, respectively, which are relatively large
values, in agreement with the analysis presented previously in \cite{pant1}.
The likelihood reaches its maximum value with $\chi_{SN+CC}^{2}\left(
w_{+}^{IV}\right)  =1419.7$ and$~\chi_{SN+CC}^{2}\left(  w_{+}^{IV}\right)
=1419.5$, resulting in $\Delta\left(  AIC\right)  =2.2$ and $2.0$ respectively.

Moreover, models $w^{V}\left(  z\right)  $ and $w^{VI}\left(  z\right)  $ fit
the data with similar values of $\chi_{SN+CC}^{2}$; however, due to the larger
number of degrees of freedom, their corresponding $\Delta\left(  AIC\right)  $
values are higher. For model $w^{VI}\left(  z\right)  $, it is important to
note that the parameter $w_{b}$ could not be constrained at all, indicating a
strong correlation between the free parameters.

These results are summarized in Table \ref{tab2A}.

According to the AIC, for the SN+CC data, the models $\Lambda$CDM and $w_{\pm
}^{IV}$ are statistically indistinguishable. Additionally, there is strong
evidence that the models $w^{III}\left(  z\right)  $, $w^{V}\left(  z\right)
$ and $w^{VI}\left(  z\right)  $ do not fit the data as well as $\Lambda$CDM.

The key conclusion from this analysis is that this dataset alone cannot
effectively distinguish between the models. This implies that, near the
present time, all asymptotic-like behaviors remain viable. However, in this
study, we utilized SN data with $z>1$, meaning that these models are not
merely considered as asymptotic solutions but rather as representations of the
general behavior of the equation of state parameter, as dictated by a linear
second-order differential equation.

%TCIMACRO{\TeXButton{B}{\begin{table}[tbp] \centering}}%
%BeginExpansion
\begin{table}[tbp] \centering
%EndExpansion
\caption{Constraints for the $SN+CC$ data.}%
\begin{tabular}
[c]{ccccccc}\hline\hline
\textbf{Model} & $\mathbf{H}_{0}$ & $\mathbf{\Omega}_{m0}$ & $\mathbf{w}_{a}$
& $\mathbf{w}_{b}$ & $\chi_{\min}^{2}$ & $\mathbf{AIC-AIC}_{\Lambda}$\\\hline
$\Lambda$CDM & $67.5_{-1.9}^{+1.6}$ & $0.331^{+0.017}$ & $\nexists$ &
$\nexists$ & $1419.5$ & $0$\\
$w_{+}^{I}\left(  z\right)  $ & $67.5_{-1.7}^{+1.7}$ & $0.337_{-0.025}%
^{+0.049}$ & $-0.24_{-0.66}^{+0.76}$ & $\nexists$ & $1419.8$ & $2.3$\\
$w_{-}^{I}\left(  z\right)  $ & $67.7_{-1.8}^{+1.8}$ & $0.324_{-0.032}%
^{+0.063}$ & $-0.05_{-0.74}^{+0.74}$ & $\nexists$ & $1419.9$ & $2.4$\\
$w^{II}\left(  z\right)  $ & $67.5_{-1.8}^{+1.8}$ & $0.345_{-0.038}^{+0.052}$
& $-0.37_{-0.65}^{+1.1}$ & $\nexists$ & $1419.9$ & $2.4$\\
$w^{III}\left(  z\right)  $ & $67.7_{-1.9}^{+1.7}$ & $0.327_{-0.029}^{+0.029}$
& $<-12.6$ & $1.1_{-2.6}^{+2.6}$ & $1419.7$ & $4.2$\\
$w_{+}^{IV}\left(  z\right)  $ & $67.6_{-1.8}^{+1.8}$ & $0.328_{-0.017}%
^{+0.022}$ & $>6.6$ & $\nexists$ & $1419.7$ & $2.2$\\
$w_{-}^{IV}\left(  z\right)  $ & $67.7_{-1.7}^{+1.7}$ & $0.333_{-0.020}%
^{+0.018}$ & $10.1_{-8.3}^{+6.4}$ & $\nexists$ & $1419.5$ & $2.0$\\
$w^{V}\left(  z\right)  $ & $67.7_{-1.7}^{+1.7}$ & $0.331_{-0.024}^{+0.024}$ &
$>5.63$ & $>4.92$ & $1419.5$ & $4.0$\\
$w^{VI}\left(  z\right)  $ & $67.8_{-1.7}^{+1.7}$ & $0.326_{-0.010}^{+0.019}$
& $15.7_{-8.3}^{+8.3}$ & $-$ & $1419.2$ & $3.7$\\\hline\hline
\end{tabular}
\label{tab2A}%
%TCIMACRO{\TeXButton{E}{\end{table}}}%
%BeginExpansion
\end{table}%
%EndExpansion

\subsection{DESI$\ $DR1: $SN+CC+BAO_{1}$ data}

We perform a constraint on these models by additionally incorporating
BAO$_{1}$ data. The maximum likelihood value is again provided by the
$\Lambda$CDM model, with $\chi_{\min}^{2}\left(  \Lambda\right)  =1435.2$. The
models $w_{\pm}^{I}\left(  z\right)  $ fit the data with $\chi_{\min}%
^{2}\left(  w_{+}^{I}\right)  =1435.8$ and $\chi_{\min}^{2}\left(  w_{-}%
^{I}\right)  =1435.4$, respectively. However, due to the different number of
free parameters, the corresponding $\Delta\left(  AIC\right)  $ values are
$2.6$ and $2.2$, respectively.

On the other hand, model $w^{II}\left(  z\right)  $ fits the data with the
same maximum likelihood value, that is, $\chi_{\min}^{2}\left(  w^{II}\right)
=1435.4$, leading to $\Delta\left(  AIC\right)  =2.2$. Moreover, for
$w^{III}\left(  z\right)  $, we find $\chi_{\min}^{2}\left(  w^{III}\right)
=1433.5$, with $\Delta\left(  AIC\right)  =2.3$.

For all these models, the best-fit parameters remain within $1\sigma$ in
comparison with the previous test without BAO data. A particularly interesting
result is that, when BAO data are included, model $w^{III}\left(  z\right)  $
has a lower $\Delta\left(  AIC\right)  $ value, contrary to the results from
the SN+CC dataset. Meanwhile, model $w_{+}^{II}\left(  z\right)  $ provides
the same $\Delta\left(  AIC\right)  $ value as in the previous test.

We continue with the second family of models, which follows from the
logarithmic parametrization. These models do not fit the cosmological data in
the same way. The corresponding values for $\chi_{\min}^{2}$ are $\chi_{\min
}^{2}\left(  w_{+}^{IV}\right)  =1435.3$, $\chi_{\min}^{2}\left(  w_{-}%
^{IV}\right)  =1434.8$, $\chi_{\min}^{2}\left(  w^{V}\right)  =1435.3$ and
$\chi_{\min}^{2}\left(  w^{VI}\right)  =1435.0$.~Thus, for the models $w_{\pm
}^{IV}$, we calculate $\Delta\left(  AIC\right)  \leq2$, indicating that these
models are statistically equivalent to the $\Lambda$CDM model. The results are
summarized in Table \ref{tab2b}.

When BAO$_{1}$ data are introduced, models $w_{\pm}^{IV}\left(  z\right)  $
and $w^{VI}\left(  z\right)  $ lead to a smaller value of $\Delta\left(
AIC\right)  $ in comparison to the $\Lambda$CDM model. However, this is not
the case for model $w^{V}\left(  z\right)  $. We note that the parameters of
model $w^{VI}\left(  z\right)  $ cannot be constrained by the data
analysis.~Models $w^{V}\left(  z\right)  $ and $w^{VI}\left(  z\right)  $ are
ruled out by these cosmological data.%

%TCIMACRO{\TeXButton{B}{\begin{table}[tbp] \centering}}%
%BeginExpansion
\begin{table}[tbp] \centering
%EndExpansion
\caption{Constraints for the $SN+CC+BAO_{1}$ data.}%
\begin{tabular}
[c]{ccccccc}\hline\hline
\textbf{Model} & $\mathbf{H}_{0}$ & $\mathbf{\Omega}_{m0}$ & $\mathbf{w}_{a}$
& $\mathbf{w}_{b}$ & $\chi_{\min}^{2}$ & $\mathbf{AIC-AIC}_{\Lambda}$\\\hline
$\Lambda$CDM & $68.8_{-1.7}^{+1.2}$ & $0.311_{-0.012}^{+0.012}$ & $\nexists$ &
$\nexists$ & $1435.2$ & $0$\\
$w_{+}^{I}\left(  z\right)  $ & $68.7_{-1.6}^{+1.2}$ & $0.303_{-0.018}%
^{+0.024}$ & $0.09_{-0.27}^{+0.39}$ & $\nexists$ & $1435.8$ & $2.6$\\
$w_{-}^{I}\left(  z\right)  $ & $68.8_{-1.6}^{+1.2}$ & $0.293_{-0.019}%
^{+0.028}$ & $-0.34_{-0.36}^{+0.47}$ & $\nexists$ & $1435.4$ & $2.2$\\
$w^{II}\left(  z\right)  $ & $68.8_{-1.6}^{+1.2}$ & $0.298_{-0.017}^{+0.029}$
& $0.20_{-0.37}^{+0.37}$ & $\nexists$ & $1435.8$ & $2.6$\\
$w^{III}\left(  z\right)  $ & $68.7_{-1.2}^{+1.2}$ & $0.307_{-0.012}^{+0.012}$
& $-17.1_{-9.0}^{+6.8}$ & $>1.57$ & $1433.5$ & $2.3$\\
$w_{+}^{IV}\left(  z\right)  $ & $68.6_{-1.5}^{+1.3}$ & $0.310_{-0.012}%
^{+0.012}$ & $\,11.3_{-4.8}^{+4.8}$ & $\nexists$ & $1435.3$ & $2$\\
$w_{-}^{IV}\left(  z\right)  $ & $68.6_{-1.4}^{+1.4}$ & $0.313_{-0.012}%
^{+0.012}$ & $9.8_{-7.7}^{+6.6}$ & $\nexists$ & $1434.8$ & $1.6$\\
$w^{V}\left(  z\right)  $ & $68.5_{-1.6}^{+1.6}$ & $0.310_{-0.014}^{+0.014}$ &
$>5.81$ & $>3.90$ & $1435.3$ & $4.1$\\
$w^{VI}\left(  z\right)  $ & $69.1_{-1.5}^{+1.2}$ & $0.312_{-0.012}^{+0.012}$
& $-$ & $-$ & $1435.0$ & $3.8$\\\hline\hline
\end{tabular}
\label{tab2b}%
%TCIMACRO{\TeXButton{E}{\end{table}}}%
%BeginExpansion
\end{table}%
%EndExpansion

\subsection{DESI$\ $DR2:$~SN+CC+BAO_{2}$ data}

We introduce the BAO$_{2}$ data as given by the DESI DR2 and we perform the
same analysis. We show that for these models there is not any difference
between the BAO$_{1}$ and BAO$_{2}$ data in terms of the $\chi_{\min}^{2}$ or
the $\Delta\left(  AIC\right)  $. $\Lambda$CDM fits the data in a better way
with the rest of the models. However, there is small statistical preference on
the $\Lambda$CDM~in comparison with the rest of the parametric dark energy models.%

%TCIMACRO{\TeXButton{B}{\begin{table}[tbp] \centering}}%
%BeginExpansion
\begin{table}[tbp] \centering
%EndExpansion
\caption{Constraints for the $SN+CC+BAO_{2}$ data.}%
\begin{tabular}
[c]{ccccccc}\hline\hline
\textbf{Model} & $\mathbf{H}_{0}$ & $\mathbf{\Omega}_{m0}$ & $\mathbf{w}_{a}$
& $\mathbf{w}_{b}$ & $\chi_{\min}^{2}$ & $\mathbf{AIC-AIC}_{\Lambda}$\\\hline
$\Lambda$CDM & $68.9_{-1.5}^{+1.2}$ & $0.311_{-0.012}^{+0.012}$ & $\nexists$ &
$\nexists$ & $1435.1$ & $0$\\
$w_{+}^{I}\left(  z\right)  $ & $68.8_{-1.5}^{+1.2}$ & $0.303_{-0.018}%
^{+0.023}$ & $0.12_{-0.28}^{+0.35}$ & $\nexists$ & $1435.7$ & $2.6$\\
$w_{-}^{I}\left(  z\right)  $ & $68.8_{-1.6}^{+1.2}$ & $0.293_{-0.020}%
^{+0.031}$ & $-0.33_{-0.42}^{+0.51}$ & $\nexists$ & $1435.8$ & $2.7$\\
$w^{II}\left(  z\right)  $ & $68.8_{-1.6}^{+1.3}$ & $0.296_{-0.019}^{+0.028}$
& $0.25_{-0.35}^{+0.35}$ & $\nexists$ & $1435.9$ & $2.8$\\
$w^{III}\left(  z\right)  $ & $68.6_{-1.4}^{+1.0}$ & $0.308_{-0.012}^{+0.012}$
& $-17.1_{-8.5}^{+7.2}$ & $2.6_{-1.5}^{+2.9}$ & $1433.4$ & $2.3$\\
$w_{+}^{IV}\left(  z\right)  $ & $68.6_{-1.5}^{+1.3}0$ & $0.310_{-0.012}%
^{+0.012}$ & $\,>7.61$ & $\nexists$ & $1435.3$ & $2.1$\\
$w_{-}^{IV}\left(  z\right)  $ & $68.8_{-1.4}^{1.4}$ & $0.313_{-0.012}%
^{+0.012}$ & $9.1_{-7.3}^{+3.1}$ & $\nexists$ & $1434.9$ & $1.8$\\
$w^{V}\left(  z\right)  $ & $68.5_{-1.7}^{+1.7}$ & $0.312_{-0.014}^{+0.014}$ &
$>6.65$ & $>4.39$ & $1435.4$ & $4.3$\\
$w^{VI}\left(  z\right)  $ & $68.9_{-1.4}^{+1.2}$ & $0.311_{-0.012}^{+0.012}$
& $14.5_{-8.5}^{+8.5}$ & $>-2.84$ & $1434.7$ & $3.6$\\\hline\hline
\end{tabular}
\label{tab2c}%
%TCIMACRO{\TeXButton{E}{\end{table}}}%
%BeginExpansion
\end{table}%
%EndExpansion

\begin{figure}[h]
\centering
\begin{subfigure}{0.65\textwidth}
\centering
\includegraphics[width=\textwidth]{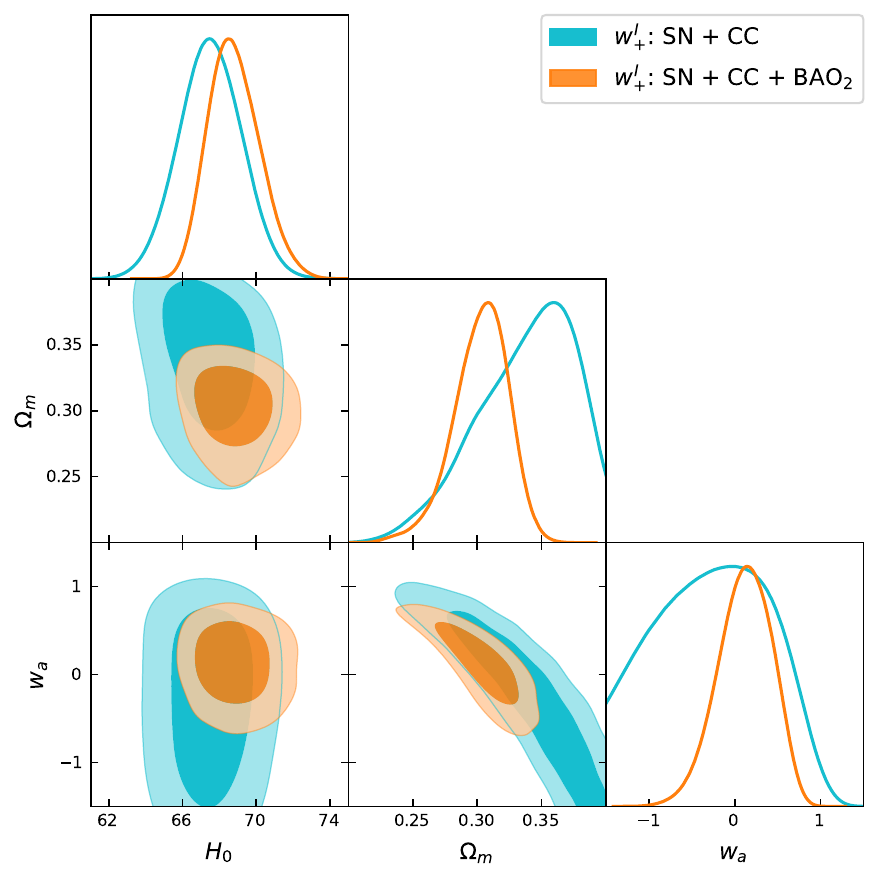}
\caption{Contour plots for $w^{I}_{+}$}
\label{fig:image11}
\end{subfigure}
\begin{subfigure}{0.65\textwidth}
\centering
\includegraphics[width=\textwidth]{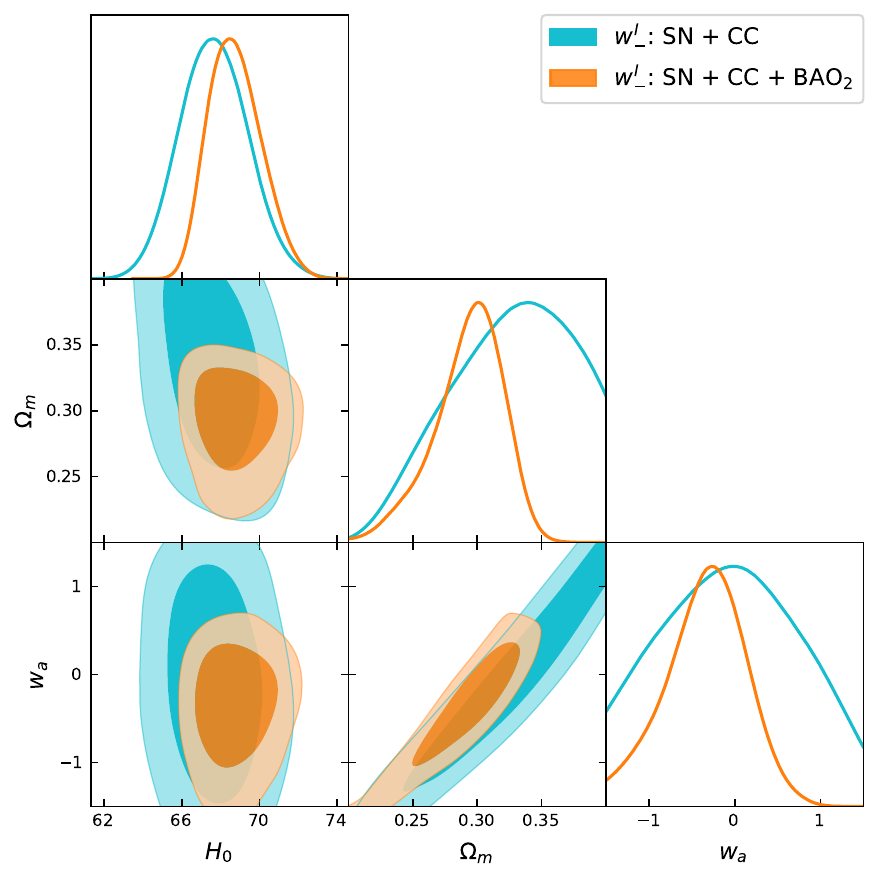}
\caption{Contour plots for $w^{I}_{-}$}
\label{fig:image22}
\end{subfigure}
\caption{Contour plots for $w_{+}^{I}$ \& $w_{-}^{I}$}%
\label{fig5A}%
\end{figure}

\begin{figure}[h]
\centering
\begin{subfigure}{0.65\textwidth}
\centering
\includegraphics[width=\textwidth]{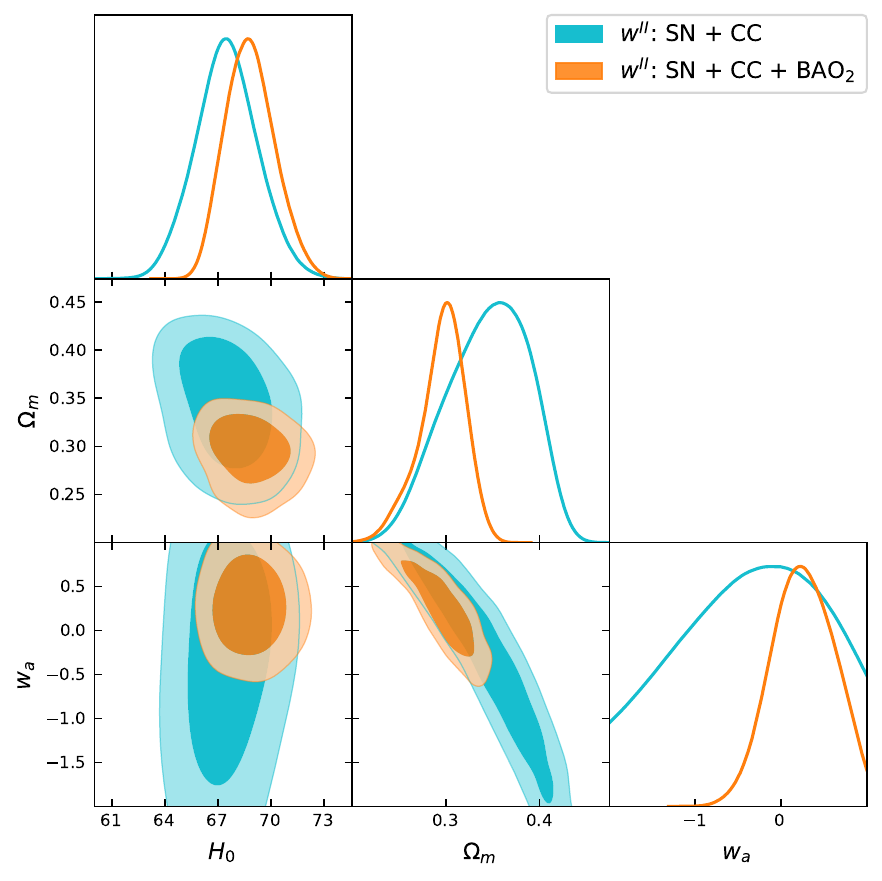}
\caption{Contour plots for $w^{II}$}
\label{fig:image33}
\end{subfigure}
\hfill\begin{subfigure}{0.65\textwidth}
\centering
\includegraphics[width=\textwidth]{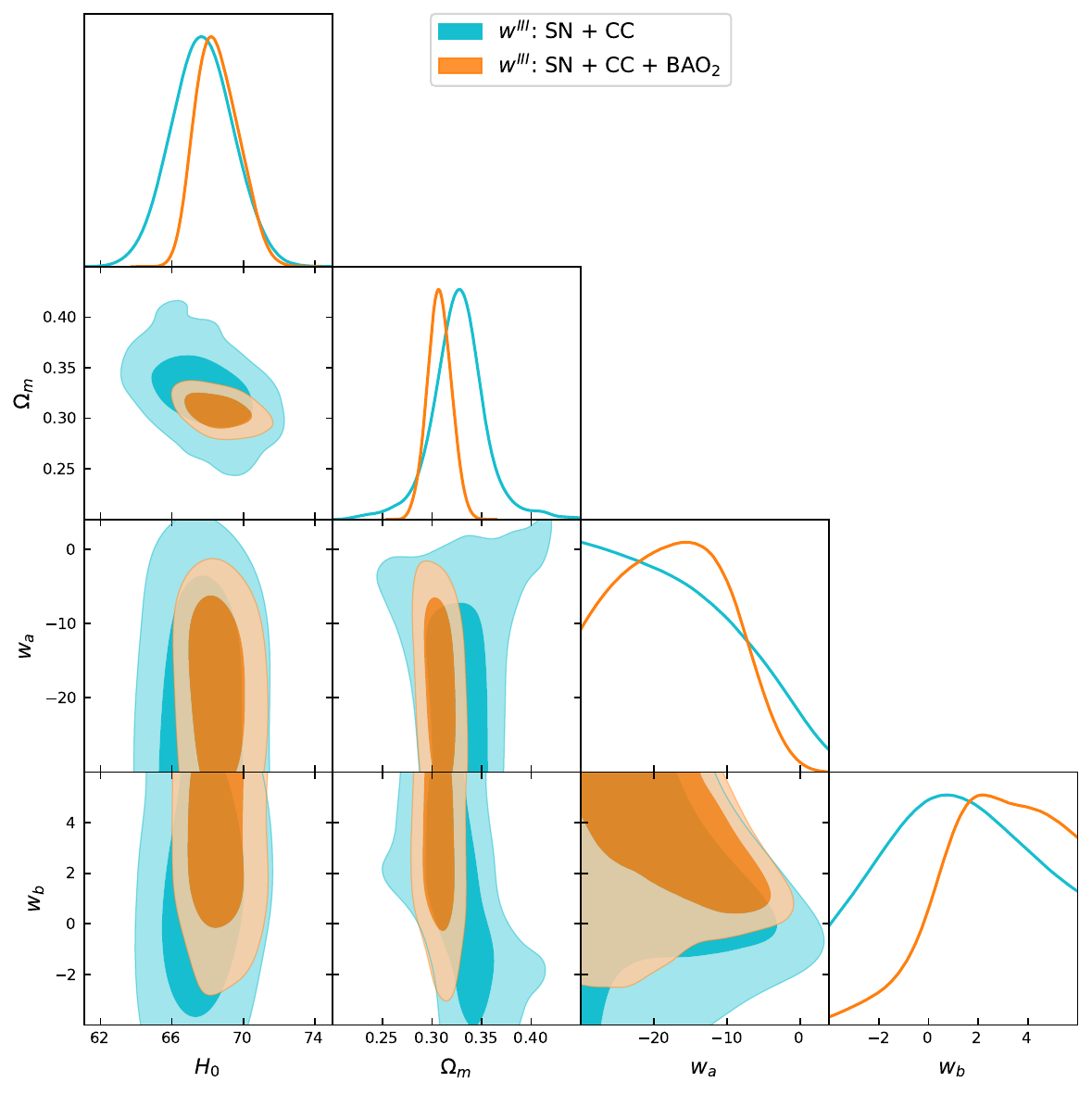}
\caption{Contour plots for $w^{III}$}
\label{fig:image23}
\end{subfigure}
\caption{Contour plots for $w^{II}$ \& $w^{III}$}%
\label{fig5B}%
\end{figure}

\begin{figure}[h]
\centering
\begin{subfigure}{0.65\textwidth}
\centering
\includegraphics[width=\textwidth]{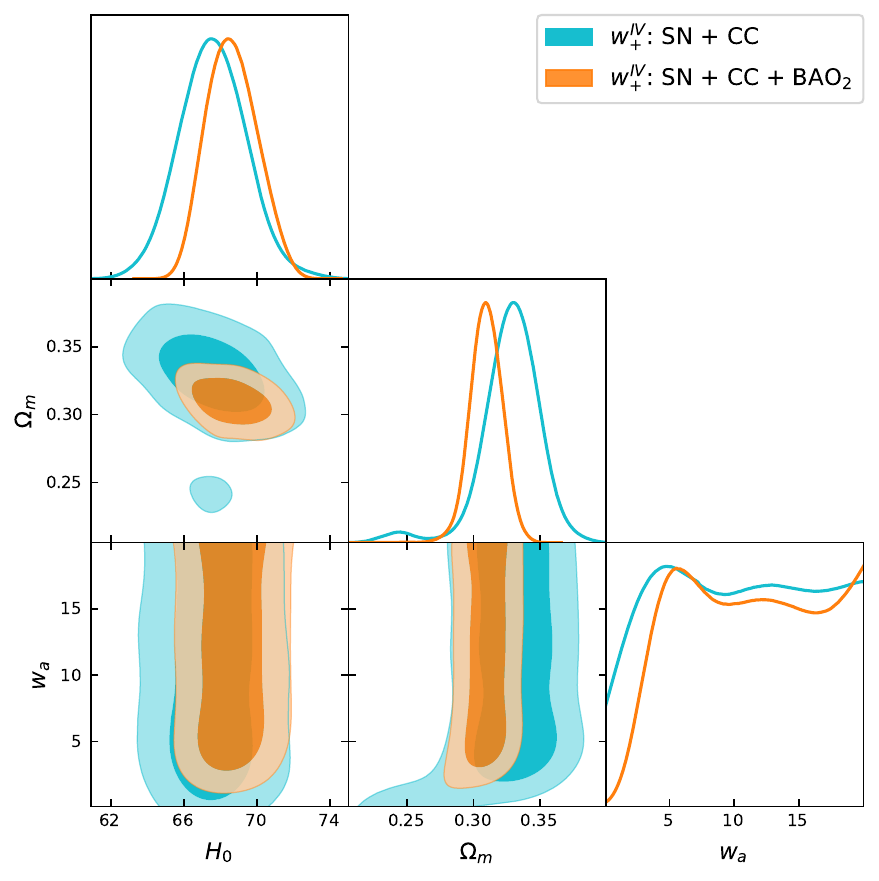}
\caption{Contour plots for $w^{IV}_{+}$}
\label{fig:image3}
\end{subfigure}
\hfill\begin{subfigure}{0.65\textwidth}
\centering
\includegraphics[width=\textwidth]{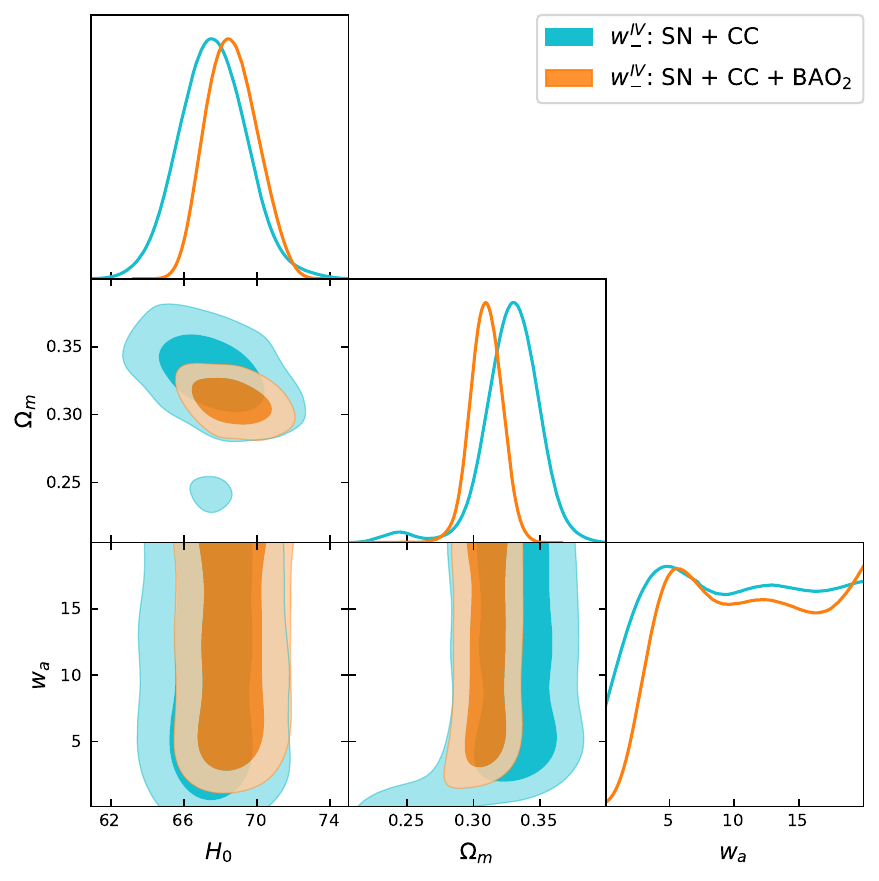}
\caption{Contour plots for $w^{IV}_{-}$}
\label{fig:image2}
\end{subfigure}
\caption{Contour plots for $w^{IV}_{+}$ \& $w^{IV}_{-}$}%
\label{fig5C}%
\end{figure}

In Figs. \ref{fig5A}, \ref{fig5B} and \ref{fig5C} we present the contour plots
for the $1\sigma$ values of the best fit parameters for models~$w_{\pm}%
^{I}\left(  z\right)  $,~$w^{II}\left(  z\right)  $,~$w^{III}\left(  z\right)
~$and$~w_{\pm}^{IV}\left(  z\right)  $ for the data sets $SN+CC$ and
$SN+CC+BAO_{2}$

\section{Conclusions}

\label{sec6}

In this work, we considered that the equation of state parameter for dark
energy is described by an arbitrary autonomous second-order differential
equation, with the cosmological constant as an equilibrium point. We derived
three functional forms for the asymptotic solutions of this differential
equation, describing the behavior of the equation of state parameter near the
equilibrium point. The free parameters of these functions depend upon the
initial value problem and the eigenvalues of the linearized differential
equation in the vicinity of the equilibrium point. Thus, constraining these
parameters provides important insights into the physical properties that the
equation of state parameter must satisfy. Based on these asymptotic solutions,
we constructed families of dynamical dark energy models.

We considered two families of models based on the parametrization of the
differential equation, each consisting of four models. The first family
follows from the linear parametrization of the equation of state parameter
with respect to the scale factor, while the second family follows from the
logarithmic parametrization of the equation of state parameter with respect to
the scale factor. By analyzing the parameter space of the free variables, we
were able to reduce the number of dependent variables to a maximum of two.
These parameters were then constrained using observational cosmological data,
specifically the Supernova, Cosmic Chronometers, and Baryon Acoustic
Oscillation datasets.

For the first family of models, which includes, $w_{\pm}^{I}\left(  z\right)
$,~$w^{II}\left(  z\right)  $ and $w^{III}\left(  z\right)  $, by using the
Supernova and Cosmic Chronometers, we found that the $\Lambda$CDM cosmology
remains the statistical preferred model. However, $w^{III}\left(  z\right)  $,
which described oscillations, has a strong degeneracy in the free parameters.
Nevertheless, when then BAO data are introduced, models ~$w^{II}\left(
z\right)  $ and $w^{III}\left(  z\right)  $ provide a better fit, although
$\Lambda$CDM~remains the preferred model.

On the other hand, for the second family of models, which follows from the
logarithmic parametrization, we found that $w_{\pm}^{IV}\left(  z\right)  $
models are statistically equivalent to $\Lambda$CDM with $\Delta\left(
AIC\right)  \leq2$. However, models $w^{V}\left(  z\right)  $ and
$w^{VI}\left(  z\right)  ~$were strongly disfavored by the data. The
parameters of model $w^{V}\left(  z\right)  ~$could not be constrained, which
means that there exists a high degree of correlation in the parameter space.%

%TCIMACRO{\TeXButton{B}{\begin{table}[tbp] \centering}}%
%BeginExpansion
\begin{table}[tbp] \centering
%EndExpansion
\caption{Constraints for the $SN+CC+BAO_{2}$ data of modified CPL models.}%
\begin{tabular}
[c]{ccccccccc}\hline\hline
\textbf{Model} & $\mathbf{H}_{0}$ & $\mathbf{\Omega}_{m0}$ & $\mathbf{w}_{0}$
& $\mathbf{w}_{a}$ & $\mathbf{w}_{b}$ & $\chi_{\min}^{2}$ & $\mathbf{AIC-AIC}%
_{CPL}$ & $\mathbf{AIC}-\mathbf{AIC}_{\Lambda}$\\\hline
CPL & $68.4_{-1.4}^{+1.1}$ & $0.308_{-0.015}^{+0.021}$ & $-0.893_{-0.056}%
^{+0.056}$ & $-0.33_{-0.48}^{+0.43}$ & $\nexists$ & $1433.3$ & $0$ & $2.2$\\
$\hat{w}_{+}^{I}\left(  z\right)  $ & $68.2_{-1.4}^{+1.2}$ & $0.301_{-0.012}%
^{+0.026}$ & $-0.894_{-0.057}^{+0.057}$ & $-0.26_{-0.74}^{+0.26}$ & $\nexists$
& $1433.4$ & $0.1$ & $2.3$\\
$\hat{w}_{-}^{I}\left(  z\right)  $ & $68.3_{-1.4}^{+1.2}$ & $0.303_{-0.016}%
^{+0.028}$ & $-0.906_{-0.060}^{+0.060}$ & $0.17_{-0.36}^{+0.65}$ & $\nexists$
& $1433.6$ & $0.3$ & $2.5$\\
$\hat{w}^{II}\left(  z\right)  $ & $68.4_{-1.4}^{+1.1}$ & $0.311_{-0.017}%
^{+0.021}$ & $-0.885_{-0.061}^{+0.061}$ & $-0.43_{-0.42}^{+0.56}$ & $\nexists$
& $1433.5$ & $0.2$ & $2.4$\\
$\hat{w}^{III}\left(  z\right)  $ & $68.3_{-1.1}^{+1.1}$ & $0.300_{-0.013}%
^{+0.013}$ & $-0.993_{-0.056}^{+0.042}$ & $<16.9$ & $0.98_{-4.3}^{+4.8}$ &
$1433.2$ & $1.9$ & $4.1$\\
$\hat{w}_{+}^{IV}\left(  z\right)  $ & $68.3_{-1.2}^{+1.2}$ & $0.295_{-0.013}%
^{+0.017}$ & $-0.914_{-0.053}^{+0.044}$ & $>7.4$ & $\nexists$ & $1433.1$ &
$-0.2$ & $2.0$\\
$\hat{w}_{-}^{IV}\left(  z\right)  $ & $68.4_{-1.2}^{+1.2}$ & $0.302_{-0.015}%
^{+0.015}$ & $-0.922_{-0.051}^{+0.040}$ & \thinspace$>6.7$ & $\nexists$ &
$1433.1$ & $-0.2$ & $2.0$\\
$\hat{w}^{V}\left(  z\right)  $ & $68.0_{-1.6}^{+1.6}$ & $0.298_{-0.015}%
^{+0.018}$ & $-0.916_{-0.053}^{+0.044}$ & $>5.6$ & $7.4_{-6.1}^{+3.0}$ &
$1433.8$ & $2.5$ & $4.7$\\
$\hat{w}^{VI}\left(  z\right)  $ & $68.4_{-1.4}^{+1.1}$ & $0.299_{-0.050}%
^{+0.042}$ & $-0.920_{-0.050}^{+0.042}$ & $16_{-11}^{+14}$ & $-$ & $1433.2$ &
$1.9$ & $4.1$\\\hline\hline
\end{tabular}
\label{tab3}%
%TCIMACRO{\TeXButton{E}{\end{table}}}%
%BeginExpansion
\end{table}%
%EndExpansion

\begin{figure}[ptbh]
\centering\includegraphics[width=0.5\textwidth]{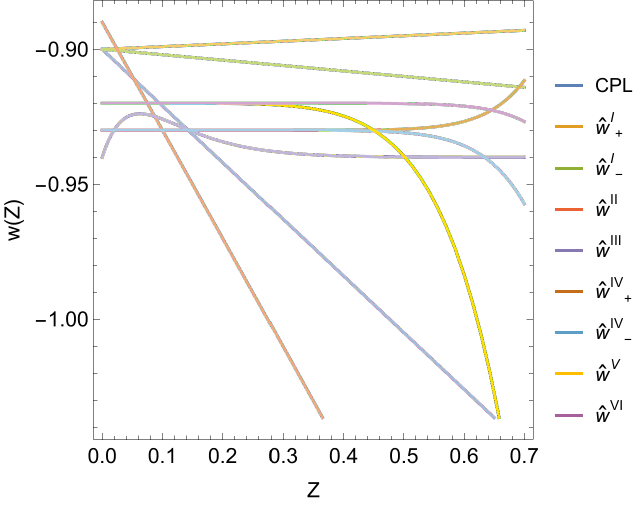}\caption{Evolution for
the dark energy equation of state parameter models: CPL, and $\hat{w}^{A}$ for
the best fit values of Table \ref{tab3}.}%
\label{fig8}%
\end{figure}

In the case for which the equilibrium point is not the $\Lambda$CDM model but
a constant value $w_{0}$, we can introduce a new class of models
\[
\hat{w}^{A}\left(  z\right)  =\left(  1+w_{0}\right)  +w^{A}\left(  z\right)
~,~
\]
with $A=I,II,III,...$, where at the present time $\hat{w}^{A}\left(
z\rightarrow0\right)  =w_{0}$. These dynamical dark energy models contain two
or three free parameters. We apply the same observational constraints as
before, but now using the CPL model as a reference. The combination of
observational data yields the best-fit parameters summarized in Table
\ref{tab3}. In Fig. \ref{fig8}, we present the evolution of the equation of
state parameters $w^{A}\left(  z\right)  $ for the best-fit parameters of
Table \ref{tab3}.

We observe that most models fit the observational data with a $\chi_{\min}%
^{2}$ value smaller than that of CPL and $\Lambda$CDM. However, due to the
different number of degrees of freedom, the application of the AIC indicates
that, in comparison to CPL, the models $\hat{w}^{A}\left(  z\right)  $ are
statistically indistinguishable. The best fit is provided by $\hat{w}_{-}%
^{IV}\left(  z\right)  $, which is known as the $n$CPL model. Finally, if when
we compare the results of Table \ref{tab2c} and \ref{tab3}, we conclude that
the DESI DR2 BAO data support models with $w_{0}>-1$.

Although these models have been introduced to describe the asymptotic behavior
of $w\left(  z\right)  $ near the equilibrium point at the present time, our
analysis shows that these models can be used for observational constraints in
the late universe. However, for CMB data, where $z\gg1$, we show that the free
parameters are degenerate, and this approximation may not be valid.

This consideration is different from the Taylor expansion of the $w\left(
a\right)  $ around the present time. Indeed, the CPL can be seen as the
first-order approximation of a nonlinear $w\left(  z\right)  $ in the Taylor
expansion. The introduction of additional terms in the Taylor expansion is
done with the charge of additional free parameters. This can lead to new
problems as discussed recently in \cite{bb11}. However, in our approach we can
reconstructing the asymptotic behavior of $w\left(  z\right)  $ at different
epochs. This can lead to further constrain\ to the physical law that $w\left(
z\right)  $ satisfies. In turn, can provide important insights into the nature
of dark energy. From this work it follows that the logarithmic parametrization
leads to parametric dark energy models which fit the datta in a better way.
Moreover, the existence of oscillations, i.e. model $\hat{w}^{VI}\left(
z\right)  $, lead to smaller values of the $\chi_{\min}^{2}$, with the cost of
a greater value for the $AIC$ due to the larger number of free parmameters. We
recall that the Chiral multi-scalar field theory \cite{ch1,ch2,ch3,ch4}, known
as warm inflation \cite{ch5}, provide such oscillationg behaviours.

\begin{acknowledgments}
AP thanks the support of VRIDT through Resoluci\'{o}n VRIDT No. 096/2022 and
Resoluci\'{o}n VRIDT No. 098/2022. Part of this study was supported by
FONDECYT 1240514 Etapa 2025. AP\ thanks Dr. F. Anagnostopoulos for a fruitful discussion.
\end{acknowledgments}

\end{document}